\documentclass[twocolumn,twoside]{article}
\makeatletter\if@twocolumn\PassOptionsToPackage{switch}{lineno}\else\fi\makeatother

\usepackage{amsmath,amssymb,amstext,tabulary,graphicx,times,caption,fancyhdr}
\usepackage[utf8]{inputenc}
\usepackage[T1]{fontenc}
\usepackage{abstract,booktabs}
\usepackage{tabularx}
\usepackage{hyperref}
\usepackage{url,multirow,morefloats,floatflt,cancel,tfrupee}
\makeatletter

\AtBeginDocument{\@ifpackageloaded{textcomp}{}{\usepackage{textcomp}}}
\makeatother
\usepackage{colortbl}
\usepackage{xcolor}
\usepackage{pifont}
\usepackage{comment}
\usepackage{framed,enumitem} 

\usepackage[nointegrals]{wasysym}
\makeatletter

\def\mcWidth#1{\csname TY@F#1\endcsname+\tabcolsep}

\def\cAlignHack{\rightskip\@flushglue\leftskip\@flushglue\parindent\z@\parfillskip\z@skip}
\def\rAlignHack{\rightskip\z@skip\leftskip\@flushglue \parindent\z@\parfillskip\z@skip}

\@ifundefined{etal}{}{}

\usepackage{ifxetex}
\ifxetex\else\if@twocolumn\@ifpackageloaded{stfloats}{}{\usepackage{dblfloatfix}}\fi\fi

\AtBeginDocument{
\expandafter\ifx\csname eqalign\endcsname\relax
\def\eqalign#1{\null\vcenter{\def\\{\cr}\openup\jot\m@th
  \ialign{\strut$\displaystyle{##}$\hfil&$\displaystyle{{}##}$\hfil
      \crcr#1\crcr}}\,}
\fi
}

\AtBeginDocument{%
  \@ifpackageloaded{endfloat}%
   {\renewcommand\efloat@iwrite[1]{\immediate\expandafter\protected@write\csname efloat@post#1\endcsname{}}}{\newif\ifefloat@tables}%
}%

\def\BreakURLText#1{\@tfor\brk@tempa:=#1\do{\brk@tempa\hskip0pt}}
\let\lt=<
\let\gt=>
\def\processVert{\ifmmode|\else\textbar\fi}

\@ifundefined{subparagraph}{
\def\subparagraph{\@startsection{paragraph}{5}{2\parindent}{0ex plus 0.1ex minus 0.1ex}%
{0ex}{\normalfont\small\itshape}}%
}{}

\newcommand\role[1]{\unskip}
\newcommand\aucollab[1]{\unskip}
  
\@ifundefined{tsGraphicsScaleX}{\gdef\tsGraphicsScaleX{1}}{}
\@ifundefined{tsGraphicsScaleY}{\gdef\tsGraphicsScaleY{.9}}{}
\def\checkGraphicsWidth{\ifdim\Gin@nat@width>\linewidth
	\tsGraphicsScaleX\linewidth\else\Gin@nat@width\fi}

\def\checkGraphicsHeight{\ifdim\Gin@nat@height>.9\textheight
	\tsGraphicsScaleY\textheight\else\Gin@nat@height\fi}

\def\fixFloatSize#1{}
\let\ts@includegraphics\includegraphics

\def\inlinegraphic[#1]#2{{\edef\@tempa{#1}\edef\baseline@shift{\ifx\@tempa\@empty0\else#1\fi}\edef\tempZ{\the\numexpr(\numexpr(\baseline@shift*\f@size/100))}\protect\raisebox{\tempZ pt}{\ts@includegraphics{#2}}}}

\AtBeginDocument{\def\includegraphics{\@ifnextchar[{\ts@includegraphics}{\ts@includegraphics[width=\checkGraphicsWidth,height=\checkGraphicsHeight,keepaspectratio]}}}

\DeclareMathAlphabet{\mathpzc}{OT1}{pzc}{m}{it}

\def\URL#1#2{\@ifundefined{href}{#2}{\href{#1}{#2}}}

\def\UrlOrds{\do\*\do\-\do\~\do\'\do\"\do\-}%
\g@addto@macro{\UrlBreaks}{\UrlOrds}

\edef\fntEncoding{\f@encoding}

\makeatother

\newif\ifmultipleabstract\multipleabstractfalse%
\usepackage[paperheight=11.02in,paperwidth=8.27in,margin=1.5cm,headsep=.5cm,bottom=1.9cm,top=2.5cm]{geometry}
\usepackage[hang,flushmargin]{footmisc}

 \date{}
\makeatletter\def\bmjIndent{1pt}
\let\@journalTitle\@empty%
\def\journalTitle#1{\gdef\@journalTitle{#1}}

\def\author#1{\gdef\@author{\hskip-\dimexpr(\tabcolsep)\hskip\bmjIndent\parbox{\dimexpr\textwidth-\bmjIndent}{\raggedright\sffamily#1}}}

\def\title#1{\gdef\@title{\raggedright\bfseries\ifx\@articleType\@empty\else\@articleType\\\fi\sffamily#1}}

\let\@articleType\@empty \def\articletype#1{\gdef\@articleType{{\normalfont\itshape#1}}}

\fancypagestyle{headings}{\fancyhf{}\fancyhead[RO]{}\fancyhead[LE]{\ifx\#journalTitle\@empty\else\sffamily\footnotesize\@journalTitle\fi}\fancyfoot[RO,LE]{\sffamily\thepage}\fancyfoot[LO,RE]{\sffamily\footnotesize\RunningAuthor}}
\fancypagestyle{plain}{\fancyhf{}\fancyfoot[R]{\sffamily\thepage}\fancyfoot[C]{\sffamily\footnotesize\RunningAuthor}}

\def\NormalBaseline{\def\baselinestretch{1.1}}

\usepackage[noindentafter]{titlesec}
\titleformat{\section}[hang]{\NormalBaseline\filright\large\bfseries\sffamily}
{\large\thesection}
{10pt}
{\noindent\MakeUppercase}
[]
\titleformat{\subsection}[hang]{\NormalBaseline\filright\bfseries\sffamily}
{\thesubsection}
{10pt}
{}
[]
\titleformat{\subsubsection}[hang]{\NormalBaseline\filright\sffamily}
{\upshape\thesubsubsection}
{10pt}
{}
[]
\titleformat{\paragraph}[runin]{\NormalBaseline\filright\itshape\sffamily}
{\theparagraph}
{10pt}
{}
[]
\titleformat{\subparagraph}[runin]{\NormalBaseline\filright\sffamily}
{\thesubparagraph}
{10pt}
{}
[]
\renewcommand\footnotemark{}
\titlespacing{\section}{0pt}{1.5\baselineskip}{.2\baselineskip}
\titlespacing{\subsection}{0pt}{1.5\baselineskip}{.2\baselineskip}
\titlespacing{\subsubsection}{0pt}{1.5\baselineskip}{.2\baselineskip}
\titlespacing{\paragraph}{0pt}{.5\baselineskip}{10pt}
\titlespacing{\subparagraph}{0pt}{.5\baselineskip}{10pt}

\makeatother

\usepackage[superscript,biblabel]{}

\def\truncatedAt{1000}
\def\typesetArtId{8974c6f9-8b6a-4cf3-b990-fabdf021eef8}    
    \makeatletter
    \@ifpackageloaded{hyperref}{}{\usepackage[pdfborder={0 0 0},bookmarks=false]{hyperref}}
    \@ifpackageloaded{hyperref}{}{}%
    \def\putUpgradeInfoBox{\@ifundefined{truncatedAt}{\def\truncatedAt{1000}}{}
    \def\up@width@one{\if@twocolumn .95\columnwidth\else .65\columnwidth\fi}%
    \def\up@width@two{\if@twocolumn .5\columnwidth\else .3\columnwidth\fi}%
    \vskip 2pc\nopagebreak
    \noindent\vbox{\centering%
    {\fontfamily{phv}\selectfont\footnotesize\color{blue}%
    	\ifx\typesetArtId\empty%
      	\href{https://typeset.io/documents}{\underline{Edit this article on \smash{Typeset}}}%
      \else%
	    	\href{https://typeset.io/edit/\typesetArtId}{\underline{Edit this article on \smash{Typeset}}}%
      \fi%
      }\\[6pt]
    \par\href{https://www.typeset.io/pricing/?source=upgrade-from-pdf}{\includegraphics[width=\up@width@one]{upgrade-logo-11Nov19.png}}\\[2pt]%
    \par%
    \fontfamily{phv}\selectfont\footnotesize\color{blue}%
    \href{https://typeset.io/}{\underline{www.\smash{typeset}.io}}%
    ~~\,\textcolor{black}{\vrule height 8pt width .7pt}~\,~%
    \href{https://www.typeset.io/orders/coupon/ZfAB8STU/?source=pricing-page-student-discount}{\underline{\smash{Looking for a Discount?}}}%
    }%
    }
    \makeatother
    
\begin{document}

\thispagestyle{plain}

\title{End-2-End COVID-19 Detection from Breath \& Cough Audio}
\author{Harry Coppock*$^1$,
            Alexander Gaskell*$^1$,
            Panagiotis Tzirakis$^1$, Alice Baird$^2$, Lyn Jones$^3$, Björn W.\ Schuller$^{1,2}$\\[-3pt]\normalsize\normalfont ~\\
            {\sffamily $^1$GLAM -- Group on Language, Audio, \& Music, 
                       Imperial College London, 
                       London, UK\\ 
                       $^2$Chair of Embedded Intelligence for Health Care and Wellbeing, University of Augsburg, Germany\\
                       $^3$Radiology Department, North Bristol NHS Trust, Bristol, UK}
       }
\def\RunningAuthor{Harry Coppock et al.}
{\thanks{*Equal contribution}}

\maketitle 

\def\keywordstitle{Keywords} 

\section*{Summary box} 
 Our main contributions are as follows:

 \begin{itemize}
     \item We demonstrate the first attempt to diagnose COVID-19 using end-to-end deep learning from a crowd-sourced dataset of audio samples, achieving ROC-AUC of \textbf{0.846}
    \item Our model, the \textbf{C}OVID-19 \textbf{Ide}ntification \textbf{R}esNet, (\textbf{CIdeR}), has potential for rapid scalability, minimal cost, and improving performance as more data becomes available. This could enable regular COVID-19 testing at a population scale
    \item We introduce a novel modelling strategy using a custom deep neural network to diagnose COVID-19 from a joint breath and cough representation
    \item We release our four stratified folds for cross parameter optimisation and validation on a standard public corpus and details on the models for reproducibility and future reference
 \end{itemize}

\noindent\hrulefill 
    
\section*{Introduction} 

The Coronavirus disease 2019 (COVID-19), caused by the severe-acute-respiratory-syndrome-coronavirus 2 (SARS-CoV-2), is the first global pandemic of the 21st century. Since its emergence in December 2019, it has led to over 75 million confirmed cases and more than 1.6 million deaths in over 200 countries (WHO)~\footnote{https://www.who.int/emergencies/diseases/novel-coronavirus-2019}. SARS-CoV-2 causes either asymptomatic infection or clinical disease, which ranges from mild to life-threatening \cite{Polidoro}. Developing a swift and accurate test, able to identify both symptomatic and asymptomatic cases, is therefore essential for pandemic control.

Vocal biomarkers of SARS-CoV-2 infection have been described, thought to relate to the clinical and subclinical effects of the virus on the lower respiratory tract, neuro-muscular function, senses of taste and smell and on proprioceptive feedback. Together, these produce a reduction in complexity of the co-ordination of respiratory and laryngeal motion in both symptomatic and asymptomatic individuals~\cite{Quatieri20}.

Recently, several audio applications have been released that capture the breath or cough of individuals. Examples include the ‘Coughvid’~\cite{orlandic2020coughvid}, ‘Breath for Science’~\footnote{https://www.breatheforscience.com}, ‘Coswara’~\cite{SharmaKKRCRGG20}, and ‘CoughAgainstCovid’~\cite{bagad2020cough}. With the release of these datasets, several studies have been published that leverage breath and/or cough signals alongside machine learning to detect the virus~\cite{brown2020exploring, bartlpokorny2020voice, ritwik2020covid, laguarta2020covid, pinkas2020sars, imran2020ai4covid}. However, these approaches try to compute representations of the breath and cough signals separately. In contrast, our approach computes a joint representation using a single model. 


We postulate that end-to-end deep learning using convolutional neural networks (CNNs) could be successfully applied to this assessment task. This article describes a proof of concept study of automatic symptomatic and asymptomatic COVID-19 recognition using combined breathing and coughing  information from audio recordings using an end-to-end CNN design. 
The code for our experiments and all details for reproduction of findings can be found at \url{https://github.com/glam-imperial/cider}. 

\begin{figure*}[ht]
\centering
    \includegraphics[width=14cm, height=5cm]{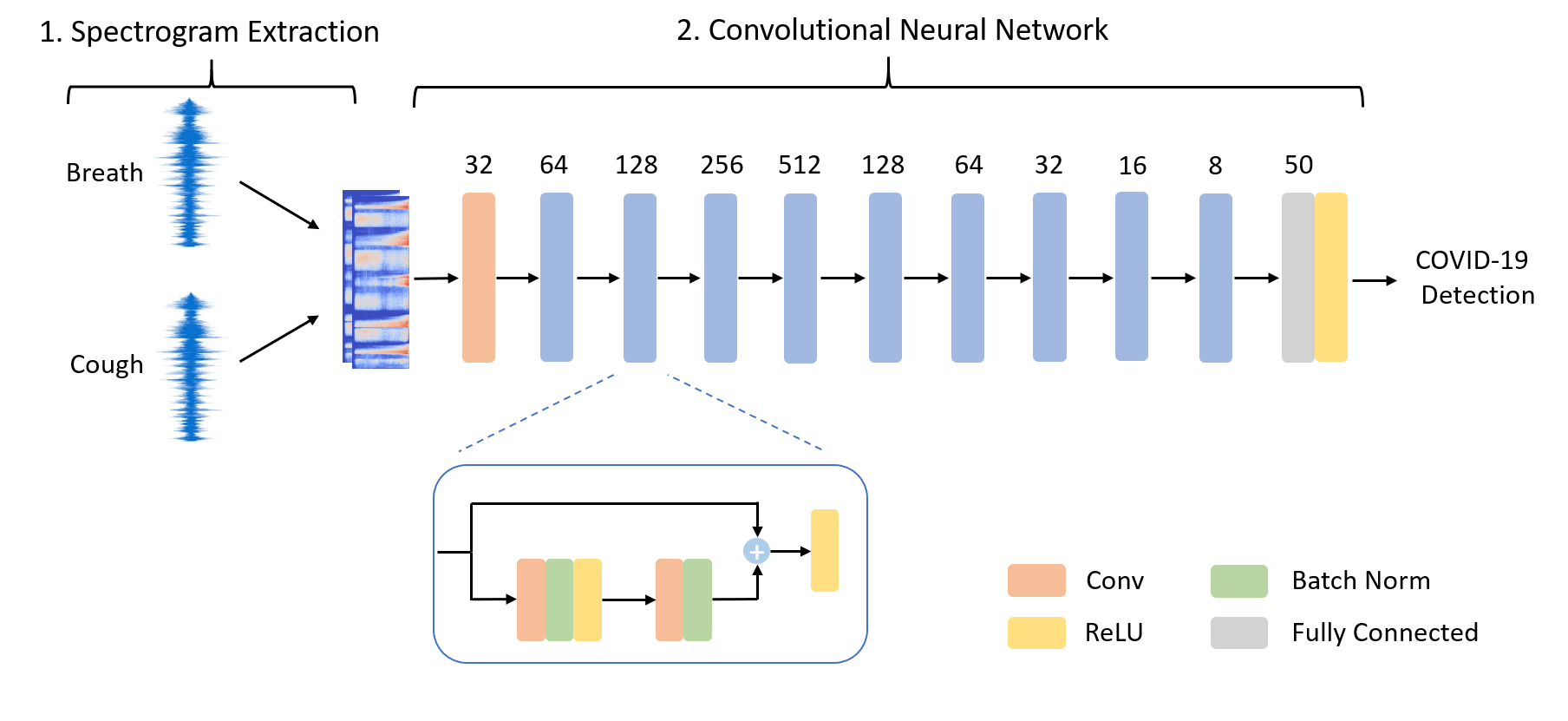}\caption{A schematic of the \textbf{C}OVID-19 \textbf{Ide}ntification \textbf{R}esNet, (\textbf{CIdeR}). The figure shows a blow-up of a residual block, consisting of convolutional, batch normalisation, and Rectified Linear Unit (ReLU) layers.}
\label{network_architecture}
\end{figure*}


\section*{Methods}
    The objective is supervised learning binary classification for diagnosing COVID-19 as positive or negative using audio signals. Our implementation, displayed in Figure~\ref{network_architecture}, has two distinct stages which are outlined below:
    
    
    \paragraph{1. Spectrogram extraction} As shown in Figure~\ref{network_architecture}, each participant in the study carried out by the University of Cambridge \cite{brown2020exploring} could submit waveform audio (WAV) files including a breath sample and a cough sample\footnote{Please see below and at https://www.covid-19-sounds.org/en/app/ for further details.}. We first compute the spectrogram of each of these WAV files to obtain a visual representation of the spectrum of audio frequencies against time. Next, we perform a log transformation, converting the spectrogram from an amplitude representation to a decibel representation. These transformations are implemented using the \textit{librosa} \cite{librosa} python package.
    
    Each WAV file lasts between one and fourty-eight seconds with a mean of ten seconds. As uniform duration is required for CNN input, we chunk the whole WAV file into $s$-second segments, using right padding for files shorter than $s$-seconds. This creates an image of size $\{F, W\}$, where $F \propto fft_n$ and $W \propto sr*s$ and $fft_n\ \text{and}\ sr$ are parameters used when computing the spectrogram. During model training, we only process one WAV segment (sampled uniformly). At inference time, we perform majority voting, whereby each chunk is processed in parallel, and the output label becomes the modal classification from each of the chunks\footnote{The mean of the output logits is taken in the case of a tied vote.}.
    
    \paragraph{2. Convolutional Neural Network} CIdeR is based on ResNets \cite{resnet}, a variant of the CNN architecture, which uses \textit{residual} blocks. As shown in Figure 1, a residual block consists of two convolutions, batch normalisation \cite{batch_norm}, and a Rectified Linear Unit (ReLU) non-linearity. These blocks use ``skip'' connections which add the output from these operations to the input activations for this layer. This alleviates the vanishing gradient problem, facilitating deeper architectures with more layers, thereby permitting richer hierarchical learnt representations. 
    The number of convolutional channels for each of CIdeR's nine layers are annotated in Figure~\ref{network_architecture}.
    
    We concatenate the log spectrograms of the breath and cough samples depth-wise, creating an $\{F, W, 2\}$ matrix as the model input. The CNN outputs a single logit which is then passed through a sigmoid layer to obtain a $(0,1)$ score, representing the probability of a COVID-positive sample. A weighted binary-cross entropy loss function \cite{weighted_loss} is used during training to address the class imbalance in the dataset. 
    
    \paragraph{Training strategy} Prior work \cite{brown2020exploring} used \textit{``10-fold-like''} cross validation during training (see the paper for details). In contrast, we implement a stratified 
    3-fold cross optimisation and additional validation partitioning using 2 / 1  (rotating development + train) / 1 (always held out fixed test) folds, respectively. This is to best optimise parameters independently of the test set with a small dataset while ensuring that the test set remains a) fixed for easier comparison with other work, and b) truly blind, eliminating the possibility of CIdeR overfitting to the test set.
    Our stratified sampling methodology ensures that our folds represent disjoint sets of participants and each of the strata (next section) are approximately uniformly distributed across each fold. To enable reproducibility, the folds are fully released in the accompanying code.
    
    \paragraph{Baseline} Our approach is not directly comparable with the study from \cite{brown2020exploring} as they do not explicitly provide their folds and discard some audio samples. To this purpose, to create a performance reference for CIdeR,
    we implement a linear kernel Support Vector Machine (SVM) \cite{SVM} baseline. We extract openSMILE features \cite{opensmile} for each wavefile following the Interspeech 2016 ComParE challenge format \cite{interspeech_compare} and perform Principal Component Analysis (PCA) \cite{bishop},
    selecting the top 100 components by highest explained variance. 
    We follow the cross optimisation
    procedure outlined above, using the development set to optimise the complexity parameter \footnote{Values between $1e^{-5}$ and $1$ on a logarithmic scale.} and reporting final results using the held-out test set.

\section*{Dataset}
\label{sec:dataset}
    The dataset used in this work consists of 517 crowdsourced coughing and breathing audio recordings from 355 participants, of which 62 participants had tested positive for COVID-19 within 14 days of the recording\footnote{The dataset used in this study is a small subset of the full dataset that has been collected by the University of Cambridge, which has yet to be made fully public. As of July 2020 the full dataset totalled 10\,000 samples from roughly 7\,000 participants.}. The samples were collected via android and web apps developed by \cite{brown2020exploring} and can be found at \hyperlink{https://www.covid-19-sounds.org}{https://www.covid-19-sounds.org}. To be classified as \textit{COVID-negative}, participants had to meet a number of stringent criteria described in \cite{brown2020exploring}.
    These participants were then divided into 3 categories: those with no cough (\textit{healthy-no-symptoms}), those with a cough (\textit{healthy-with-cough}) and those who had asthma (\textit{asthma-with-cough}). The \textit{COVID-positive} class is constituted of the 62 COVID-positive participants and is further divided into the sub classes \textit{COVID-no-cough}, and \textit{COVID-cough} representing 39 COVID positive participants without a cough and 23 participants with a cough, respectively.

\section*{Experiments \& Results}

As indicated above, we perform a 3-fold cross optimisation using the rotating development plus train folds. Recall that the test set is fixed and always held out during optimisation. For evaluation metrics, we utilise the Area Under Curve of the Receiver Operating Characteristics curve (AUC-ROC), and Unweighted Average Recall (UAR), both of which are robust to imbalanced datasets. AUC-ROC maps the relationship between sensitivity and the false positive rate as the classification threshold is varied, and UAR computes the mean recall per class. The models' performance is sensitive to initialisation parameters,  so we report the mean and standard deviation from three training runs. \autoref{tab:hyp} details our hyperparameter search and optimal values used for the final model.

\begin{table}[t!]
\centering
\caption{Overview of the hyperparameter search detailing the interval, step size, and optimal parameters (used to obtain the reported figures in this article -- for details cf.\ the above named GitHub repository). Hyperparameters were optimised for task 4, and subsequently used on all tasks. $^*$Interval constructed using a logarithmic scale. Adam \cite{adam} was used for optimisation.
}
\small
\begin{tabularx}{250pt}{l r r r r}
    \toprule
    \textbf{Parameter} & \textbf{Min.} & \textbf{Max.} & \textbf{Step} & \textbf{Optimal} \\
    \midrule
     Learning rate & $5e^{-5}$ & $5e^{-4}$ & $5e^{-5}$ & $1e^{-4}$ \\
     Batch size & 8 & 32 & 2$^*$ & 16 \\
     Audio segment length [s] & 1 & 8 & 2$^*$ & 8 \\
     Spectral bands (fft$_n$) [\#]  & 512 & 2\,048 & 2$^*$ & 1\,024 \\
     Sample rate $sr$ [kHz] & 24 & 48 & 2$^*$ & 24 \\
    \bottomrule
\end{tabularx}
\label{tab:hyp}
\end{table}%

Our model performs the three tasks described in the dataset publication \cite{brown2020exploring}, and an additional fourth task. 
Tasks 1-4 are as follows:

\paragraph{Task 1} Distinguishing between \textit{COVID-positive} and the strata \textit{healthy-no-symptoms} (62 vs 245 participants).
\paragraph{Task 2} Distinguishing between \textit{COVID-positive} participants with a cough (\textit{COVID-cough}) and the strata \textit{healthy-with-cough} (23 vs 30 participants).
\paragraph{Task 3} Distinguishing between \textit{COVID-positive} participants with a cough (\textit{COVID-cough}) and the strata \textit{asthma-with-cough} (23 vs 19 participants).
\paragraph{Task 4} Distinguishing between \textit{COVID-positive} and \textit{COVID-negative} (62 vs 293 participants).\\

\noindent
Note that the number of participants deviates from \cite{brown2020exploring}, as we also use those audio clips shorter than two seconds resulting in partially more participants considered.

Results obtained for each task are shown in Table~\ref{tab:exp}, alongside the baseline. CIdeR outperforms on all tasks bar task 2 with high margin on both metrics. The results for tasks 1, 3, and 4 are statistically significant with a level of significance of 0.01 in a two-sided two sample t-test for difference in sample means.

\begin{table}[t!]
\centering
\caption{Results of the models on tasks 1-4 for 3-fold optimisation of the number of training epochs based on the rotated development sets using the frozen optimal model parameters from Table~\ref{tab:hyp}. [Train+development / test] sample counts are displayed alongside the task. Testing is performed on the held out test fold, each. The mean Area Under Curve of the Receiver Operating Characteristics curve (AUC(-ROC)) and the Unweighted Average Recall (UAR) are displayed. A 95\% confidence interval is also shown following \cite{hanley_mcneil_1982} and the normal approximation method for AUC-ROC and UAR respectively. Scores in bold indicate  significant results with $\alpha = 0.05$ using a 2-sample t-test for no difference in means between the baseline and CIdeR based on the standard deviation from the 3-fold cross validation.
}
\small
\begin{tabularx}{250pt}{l|X X|X X}
    \toprule
    \textbf{Task} & \multicolumn{2}{c}{\textbf{CIdeR}} & \multicolumn{2}{c}{\textbf{Baseline}}\\
    \cline{2-5}
     & \multicolumn{1}{c}{AUC} & \multicolumn{1}{c}{UAR} & \multicolumn{1}{c}{AUC} & \multicolumn{1}{c}{UAR} \\
    \midrule

     1\ \ [$688\ /\ 238$] & \textbf{.827}$\pm$.051 & .770$\pm$.053 & .697$\pm$.066 & .677$\pm$.059 \\
     2\ \ [$146\ /\ 28$]* & .570$\pm$.216 & .535$\pm$.185 & .628$\pm$.208 & .583$\pm$.183 \\
     3\ \ [$118\ /\ 32$]* & \textbf{.909}$\pm$.130 & \textbf{.774}$\pm$.145 & .559$\pm$.220 & .506$\pm$.173\\
     4\ \ [$684\ /\ 350$] & \textbf{.846}$\pm$.040 & \textbf{.765}$\pm$.044 & .721$\pm$.053 & .654$\pm$.050 \\

    \bottomrule
\end{tabularx}
\label{tab:exp}

\footnotesize{*It is questionable whether the normality assumption holds at these small sample sizes. The confidence interval estimates should therefore be taken lightly.}
\end{table}%


\section*{Discussion}

    The results in \autoref{tab:exp} demonstrate two key points: 1) it is possible to diagnose COVID-19 using a CNN-based model trained on crowdsourced data; 2) CIdeR obtains a high AUC-ROC of 0.846 on task 4, the task which uses the entire sample, and so represents the most pertinent task. These suggest that jointly processing breath and cough audio signals using a CNN-based classifier could act as an effective and scalable method for COVID-19 diagnosis. 
    
    The only task where CIdeR fails to outperform the baseline in our experiments is task 2. We posit this is jointly due to the small number of samples and the similarity of audio patterns of healthy participants with a cough and those with COVID-19, creating a challenging task. We leave further analysis for future work.
    
    A key limitation of this study is the size and demographics of the publicly available dataset \cite{brown2020exploring}. We are limited to 62 COVID-positive participants, limiting the breadth of any conclusions we draw. Our control group, COVID-free participants, is not a random sample as participation required the subject lived in a country with low COVID-19 rates, among other criteria.
    How representative these audio biomarker features are of the wider population is therefore still an open question.
    Importantly, before a such a technology can be deployed, evaluation on a larger, more representative dataset is necessary. As alluded to in \cite{mckimm_2020}, pandemics have historically led to breakthroughs in healthcare. If AI-driven screening is to be one of these breakthroughs from the 2020 COVID-19 pandemic, a more comprehensive dataset is required.

\section*{Conclusion}
    Wholesale testing of the population is a promising avenue for identifying and controlling the spread of COVID-19. A digital audio-collection and diagnostic system could be deployed to the majority of the population and performed daily at minimial cost, e.\,g., for pre-selection for more reliable diagnoses or monitoring of spread, and therefore holds great potential. This study introduced the \textbf{C}OVID-19 \textbf{Ide}ntification \textbf{R}esNet, (\textbf{CIdeR}), which demonstrated a strong proof-of-concept for applying end-to-end deep learning to jointly learning representations from breath and cough audio samples. 
    This was despite a small dataset; given more samples, it seems likely that CIdeR's diagnostic capabilities would significantly increase.
    
\section{Acknowledgements}

    The authors give their thanks to the help provided by their colleagues Mina A.\ Nessiem and Mostafa M.\ Mohamed.
    
    The University of Cambridge does not bear any responsibility for the analysis or interpretation of the data used herein, which represents the own view of the authors of this communication.

\bibliography{refs}
\bibliographystyle{unsrt}


    
\end{document}